# Achromatic Metalens in the Visible and Metalens with Reverse Chromatic Dispersion

M. Khorasaninejad[1*], Z. Shi[2*], A. Y. Zhu[1], W. T. Chen[1], V. Sanjeev[1,3], A. Zaidi[1] and F. Capasso[1]

[1]*Harvard John A. Paulson School of Engineering and Applied Sciences, Harvard University, Cambridge, Massachusetts 02138, USA*
[2]*Department of Physics, Harvard University, Cambridge, Massachusetts 02138, USA*
[3]*University of Waterloo, Waterloo, ON, N2L 3G1, Canada*

*These authors contributed equally to this work.

**In this letter, we experimentally report an achromatic metalens (AML) operating over a continuous bandwidth in the visible. This is accomplished via dispersion engineering of dielectric phase shifters: titanium dioxide nanopillars tiled on a dielectric spacer layer above a metallic mirror. The AML works in reflection mode with a focal length independent of wavelength from λ=490 nm to 550 nm. We also design a metalens with reverse chromatic dispersion, where the focal length increases as the wavelength increases, contrary to conventional diffractive lenses. The ability to engineer the chromatic dispersion of metalenses at will enables a wide variety of applications that were not previously possible. In particular, for the AML design, we envision applications such as imaging under LED illumination, fluorescence, and photoluminescence spectroscopy.**

The design principles of diffractive optics date back to microwave devices, where an array of metallic antennas[1-3] was used to control the radiation pattern of an incident electromagnetic wave. With progress in fabrication techniques, numerous planar devices, such as gratings and filters [4-9], lenses[6, 10-12], polarization-sensitive devices[13-15], and mirrors[16-19] have been realized in the infrared and visible. Metasurfaces[20-27] have added another dimension to diffractive optics by enabling control over the basic properties of light (phase, amplitude and polarization) with subwavelength spatial resolution. One particular device that has attracted a lot of attention is the planar lens[18, 28-47] (metalens). Due to their planar configuration, metalenses have potential applications in several multidisciplinary areas including imaging, spectroscopy, lithography, and laser fabrication. High numerical aperture (NA) and high efficiency metalenses have been demonstrated in both near infrared[39] and visible[40, 43]. It has also been shown that multifunctional metalenses[33, 42, 48, 49] can be achieved without adding design and fabrication complexity. Multiorder diffractive lenses[50] and multiwavelength achromatic metalenses[51-57] have been reported to compensate for chromatic dispersion at several discrete wavelengths but achieving achromatic focusing over a significant bandwidth has proven to be challenging for both Fresnel lenses and metalenses. The achromatic performance is required for a wide variety of applications, in which either the illumination source (e.g. LED imaging) or the signal (e.g. fluorescence and photoluminescence signals) has a substantial bandwidth. In a recent work, a broadband achromatic cylindrical lens in the visible was demonstrated [57]. However, it has a low numerical aperture (NA=0.013), requires three dimensional fabrication (grayscale lithography), and restricts the constituent materials to resists. Here, we demonstrate a visible wavelength achromatic metalens (AML), which requires only a one-step lithography process, with NA=0.2, and a constant focal length over a continuous range of wavelengths, from 490 nm to 550 nm.

This bandwidth, which is close to that of an LED, enables numerous applications using metalenses, which were not previously possible. In addition, we design a metalens with reverse chromatic dispersion, which further proves the effectiveness of our approach to achieve planar lenses with tailored chromatic dispersion.

**Metalens Design:** The required phase profile to achieve diffraction-limited focusing for collimated incident light is:

$$\varphi(x, y, \lambda) = C(\lambda) - \frac{2\pi}{\lambda}\left(\sqrt{x^2 + y^2 + f^2} - f\right) \quad (1)$$

where $\lambda$ is wavelength, $f$ is focal length, and $C$ is a reference phase, which is chosen to be wavelength dependent. One can see the wavelength dependence of the required phase and the challenge to achieve an achromatic lens (fixed focal length over a bandwidth). In refractive lenses, phase is achieved via volumetric propagation, and due to low dispersion of the material used (e.g. silica), chromatic aberration is not pronounced and therefore can be corrected in a straightforward manner with well-established strategies using multiple lenses. In metasurface-based diffractive lenses however, the required phase is commonly achieved via waveguiding effects[43] or effective medium designs[11], where both often have undesired dispersion. In addition, the interference of light transmitted or reflected from this phase mask (lens) is used to focus light. Contributions from both the phase realization process and the interference mechanism result in large chromatic aberrations in diffractive lenses. To understand this better, let us consider a dielectric-based metalens, where each phase shifter (PS), nanopillar, acts as a truncated waveguide with predetermined dispersion. An essential factor in designing a metalens is optimizing the PS's geometric parameters, such as the diameter, height and center-to-center distance of nanopillars[43], to achieve the required phase coverage of 0-2π. The desired phase coverage is obtained by varying a geometric parameter, generally the lateral dimension of nanopillars, from minimum to maximum. The fabrication constraints in making high aspect ratio nanopillars (small diameter, large height $H$ that is kept constant across the metalens for fabrication reasons) make it difficult to extend the phase coverage beyond $2\pi$. This basically gives only one option in selecting the required phase for a given wavelength, which translates into a specific geometric design parameter. This implies that in the selection of an appropriate PS at a design wavelength, we often cannot simultaneously satisfy the required phase at other wavelengths. Here, we break away from this constraint by designing the PS such that it not only provides multiple $2\pi$ phase coverage[50], but also anomalous dispersion – both of which offer extra degrees of freedom for designing the AML. The PS of our AML (Figure 1a) is a titanium dioxide (TiO$_2$) nanopillar (with a square cross-section) on a metallic mirror with a thin layer of silicon dioxide (spacer) in-between (Figure 1b,c). Our building block is similar to what is used in Ref. [58] where amorphous silicon nanopillars are used to design achromatic lenses in the near infrared. Here, our primary goal is to maximize the phase coverage, and to do so, one needs to optimize the nanopillars' parameters: cross-section shape (square in our case), width, center-to-center distance, and height. The advantage of using a square cross section is twofold. Firstly, it maximizes the filling factor range: from zero (no nanopillar) to ~1 (width equals to center-to-center distance), which is necessary to increase the phase coverage. Secondly, it guarantees polarization insensitive operation of the AML. The other parameter is the nanopillar width. While the smallest achievable width (here 80 nm) is limited by fabrication constraints, the largest one is less than the center-to-center distance between neighboring nanopillars. Center-to-center distance $U$ defines the sampling rate of the phase profile (Eq. (1)) and must satisfy the Nyquist sampling criterion ($U < \frac{\lambda}{2NA}$). Additionally, in order to excite guided mode resonance [59, 60] responsible for anomalous dispersion of the PSs (Supporting Information Note 1), the center-to-center distance should be larger than all wavelengths (across the design bandwidth) in the $TiO_2$. This distance however should also be smaller than all wavelengths (across the design bandwidth) in free space to suppress higher diffraction orders. Therefore, we select the center-to-center distance to be *U=480 nm*. Regarding the nanopillar, the larger the height *H*, the greater the phase coverage. Height is limited by fabrication constraints to 600 nm. Note that we design the AML to operate in reflection, which has the advantage of increasing the overall phase coverage (longer optical path).

Figure 1d shows the phase shift as function of the width of the PS at two different wavelengths (500 nm and 550 nm). Note that the phase is folded between 0 to $2\pi$. As it is clear from the figure, for each wavelength, we have several choices of widths that provide the same phase, but have different dispersive responses. In other words, at the wavelength of 500 nm, there are multiple widths that give the same phase (module of $2\pi$), but these widths provide different phases at the wavelength of 550 nm. Another important observation is that, contrary to previous metalens designs[43], the unfolded phase shift as a function of the nanopillar width does not experience a monotonic increase. As shown in Figure 1d, there are a few ranges of widths in which the phase decreases for increasing width. This anomalous behavior stems from the excitation of guided mode resonances (Supporting Information Note 1) and provides another degree of freedom to engineer the dispersive response of each PS.

We chose the bandwidth from 490 nm to 550 nm and discretized it into six equally spaced wavelengths. The required phase at each wavelength is calculated by Eq. (1), knowing the focal length, *f=485 μm*, and diameter, *D=200 μm*, of the AML. The reference phase *C* can be different for different wavelengths and thus introduces a "knob" to minimize the difference between the implemented and required phase simultaneously for all selected wavelengths. We determine $C$ by utilizing the particle swarm optimization algorithm of Matlab. Optimized solutions are shown in Figure S7, where we overlaid the required phase and the achieved phase at several wavelengths.

Next, we analyzed the performance of the AML by means of the Fresnel–Kirchhoff integral[61] (instead of full-wave analysis via FDTD), due to the large size of the AML and our limited computational resources. The focal length of an AML as a function of wavelength is calculated using this method and shown in Figure 2. We see that the focal length is nearly constant from 490 nm to 550 nm, with a fractional change of 1.2% ($\frac{\max(f) - \min(f)}{mean(f)} \times 100\%$), thus verifying the effectiveness of our chromatic dispersion engineering approach. It is notable that previous methods[51, 52, 55], which use coupled dielectric scatterers to realize multiwavelength achromatic metalenses, cannot be utilized here. This is due to their resonant design, which results in very different focal lengths at wavelengths away from the selected ones.

**Fabrication and Characterization:** To fabricate the AML, we started with a fused silica substrate. An aluminum mirror with a thickness of 110 nm (larger than the skin depth of light in the visible) was deposited using electron beam deposition. Then, a 180 nm thick silicon dioxide film was grown via plasma enhanced chemical vapor deposition. Finally, $TiO_2$ nanopillars were fabricated using the method developed in Ref. [62]. Scanning electron microscope images of the fabricated AML are shown in Figure 3a,b. Nanopillars with high aspect ratios, up to 7.5, and vertical side-walls are achieved. The optical image of the AML is also shown in Figure 3c. The AML has a diameter of 200 μm, with a focal length of 485 μm, giving a NA=0.2.

To characterize the AML, we use a custom-built optical setup (Figure 3d). First, a collimated beam is passed through a 50:50 beam splitter to illuminate the AML. Then, the focused reflected beam is collected by the imaging system via the beam splitter. The imaging system consists of two objectives and their corresponding tube lenses, providing a magnification of 40. We verified this magnification by imaging an object of known size. Due to limited space between the first objective and the AML, we chose a long working distance (WD=34 mm) objective (Mitutoyo M Plan Apo 10×, NA=0.28) and paired it with its tube lens (focal length of 200 mm). Then this image is magnified again by a 4× objective (Olympus PLN Plan Achromat 4×) paired with its tube lens (focal length of 180 mm) to form the final image on a sCMOS camera (Andor Zyla). The AML was mounted on a motorized stage (with a minimum movement step of 100 nm) to adjust the distance between the AML and first objective (10× objective). Using this setup, we moved the AML in steps of 2 μm along the z-axis and recorded the intensity profile in the *xy*-plane. By stitching these intensity profiles, we reconstructed the intensity profile of the reflected beam by the AML in the *xz*-plane. This intensity profile illustrates the propagation of reflected beam right before and after the focal distance, as shown in Figure 4a. We see that the focal length remains unchanged for wavelengths from 490 nm to 550 nm. This becomes clearer by looking at the measured focal length as a function of wavelength (Figure 2). The measured focal lengths closely resemble those predicted by the simulation, in particular across the design bandwidth. In the measured intensity profiles in the *xz*-plane (Figure 4a), one can see secondary intensity peaks before and after the focal spots, an effect which is more pronounced for wavelengths at the edge of the bandwidth. To understand their origin, we ran a simulation shown in Figure 4b. It is evident that intensities of secondary intensity peaks are significantly smaller than those observed in the experiment. This implies that fabrication imperfections enhance this effect (Figure S8), which can be mitigated by further optimizing the fabrication parameters including the electron beam lithography dosage. Secondary peaks in the simulations arise from the fact that we have a mismatch between the required and realized phase profiles (Figure S7). The latter also reduces the focusing efficiency of AML (Figure S9).

In addition, Figure 5 shows the intensity profiles of the AML in the *xy*-plane at a fixed distance of 485 μm from the AML. These intensity profiles resemble the focal spots (also shown in Figure 5, bottom row), which is consistent with the observed small change in focal lengths for the different wavelengths. The importance of dispersion engineering becomes evident when comparing these results with the measured results of a metalens based on geometric phase[40]. Since the latter is not corrected for chromatic dispersion, as shown in Figure 2, the focal length monotonically reduces as the wavelength increases. To demonstrate the effectiveness and versatility of our approach, we also designed a metalens with reverse chromatic dispersion (Figure 6), i.e. the focal length increases with wavelength.

In conclusion, we have demonstrated metalenses with tailored chromatic dispersion. We have theoretically and experimentally shown planar lenses with an achromatic response, where the focal length remains unchanged for a 60 nm bandwidth in the visible. Dispersion engineering is an important topic in optics, with a wide range of possible applications. Demonstration of this achromatic metalens, along with the design of metalenses with reverse chromatic dispersion, proves that one can break away from the constraints of conventional diffractive optics and explore opportunities for the development of new components with desired dispersion. For example, by cascading metalenses with opposite dispersions, one can potentially achieve achromatic lenses with larger bandwidths. In addition, we believe that a similar design principle can be used to realize transmissive achromatic metalenses. In this case, a higher refractive index material such as gallium phosphide can be helpful due to its larger refractive index (larger optical path resulting in higher phase coverage and more degrees of freedom in the design). In addition, by using high refractive index phase shifters, one can potentially increase the bandwidth of operation for reflective achromatic metalenses.

AUTHOR INFORMATION

Corresponding Authors

*E-mail: capasso@seas.harvard.edu.

*E-mail: khorasani@seas.harvard.edu.

The authors declare no competing financial interest.

**Acknowledgements**

This work was supported in part by the Air Force Office of Scientific Research (MURI grant# FA9550-14-1-0389 and FA9550-16-1-0156). A. Y. Z. thanks Harvard SEAS and A*STAR Singapore under the National Science Scholarship scheme. This work was performed in part at the Center for Nanoscale Systems (CNS), a member of the National Nanotechnology Coordinated Infrastructure (NNCI), which is supported by the National Science Foundation under NSF award no. 1541959. CNS is a part of Harvard University.

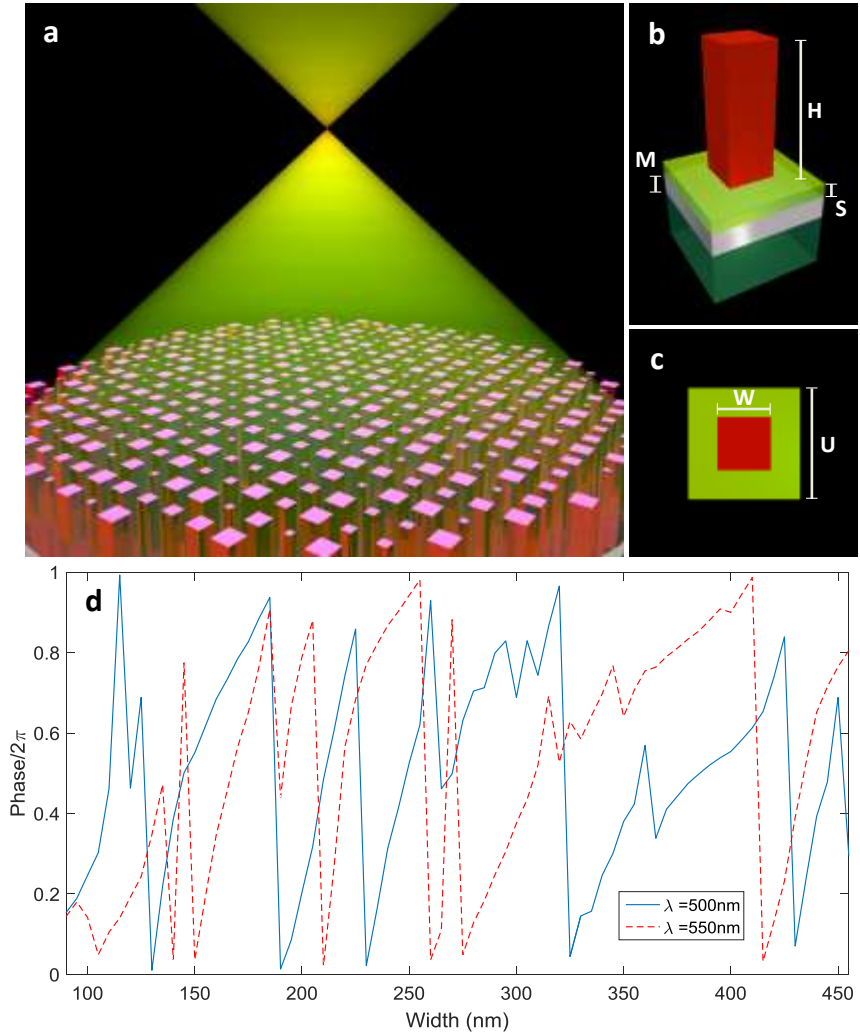

**Figure 1.** (a) Schematic of an achromatic metalens (AML). The AML focuses a collimated incident light into a spot in reflection mode. (b) The building block of the AML consists of a titanium dioxide (TiO$_2$) nanopillar, with height *H=600 nm,* on a substrate. The nanopillar has a square cross-section with width *W*. By adjusting the width, the reflection phase can be controlled. The substrate is an aluminum-coated fused silica with a thin film of silicon dioxide deposited on top. Aluminum and silicon dioxide have thicknesses *M=110 nm* and *S=180 nm*, respectively. (c) Top-view of the building block shows the width of TiO$_2$ nanopillar with unit cell size *U=480 nm*. (d) Computed reflection phase shift as a function of the nanopillar width at two different wavelengths of 500 nm and 550 nm.

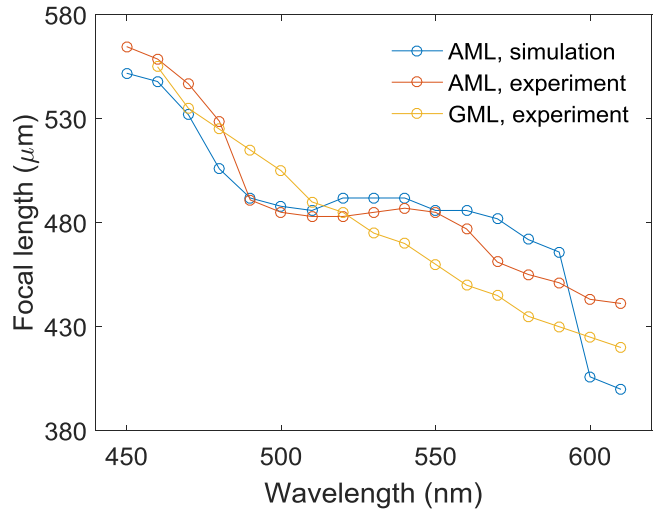

**Figure 2.** *AML, simulation:* calculated focal length of the achromatic metalens (AML) as function of wavelength. *AML, experiment:* measured focal length of the fabricated AML versus wavelength. The measured fractional change of focal length across the bandwidth (490 nm-550 nm) is 1.5% very close to that of predicted by simulation (1.2%). The AML has a diameter of 200 μm and focal length of 485 μm at λ=530 nm. *GML, experiment:* measured focal length of a fabricated geometric-phase based metalens (GML). The required phase is imparted via the rotation of titanium dioxide nanofins. The use of geometric phase has a high tolerance to fabrication errors thus GML can serve as a robust reference.

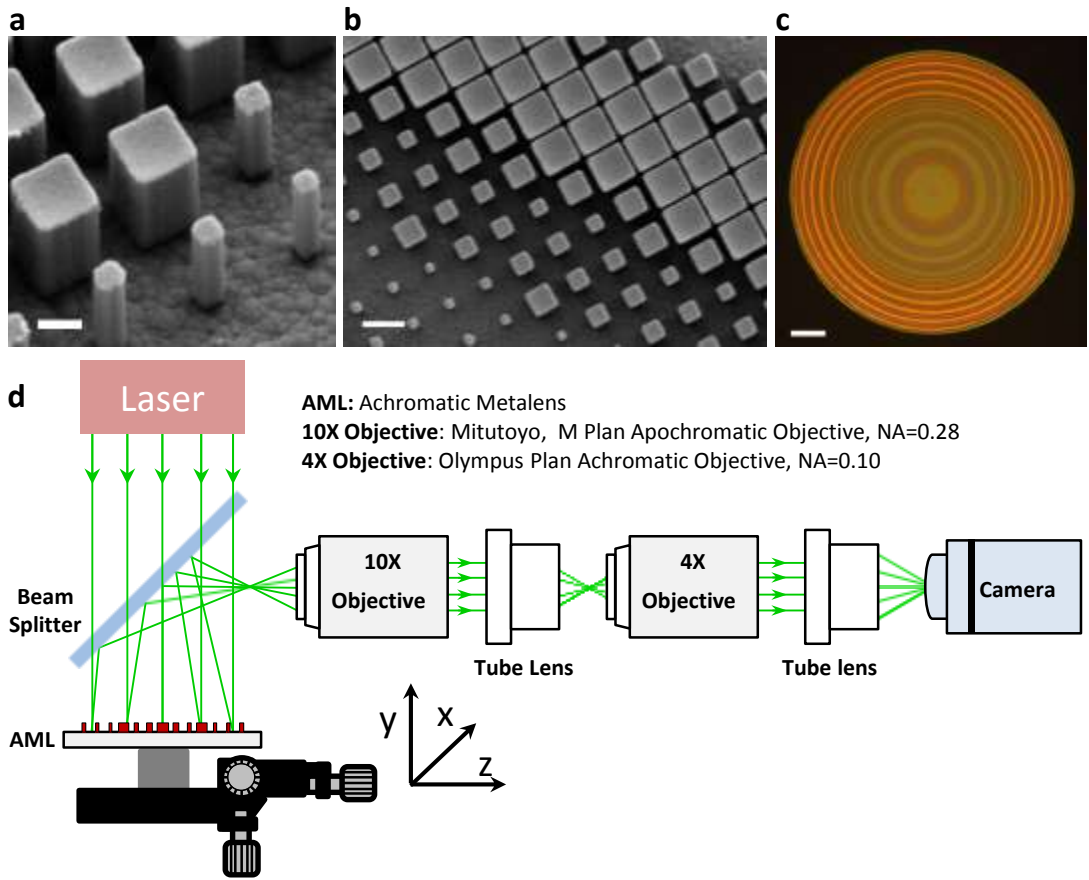

**Figure 3.** (a) Side-view scanning electron microscope (SEM) image of the fabricated achromatic metalens (AML). Scale bar: 200 nm. (b) Top-view SEM image of the AML. Scale bar: 500 nm. (c) Optical image of the AML. Scale bar: 25 μm. (d) Schematic of custom-built setup to characterize the AML. The AML reflects and focuses the collimated incident light. We cascaded two objectives to increase the imaging magnification to 40.

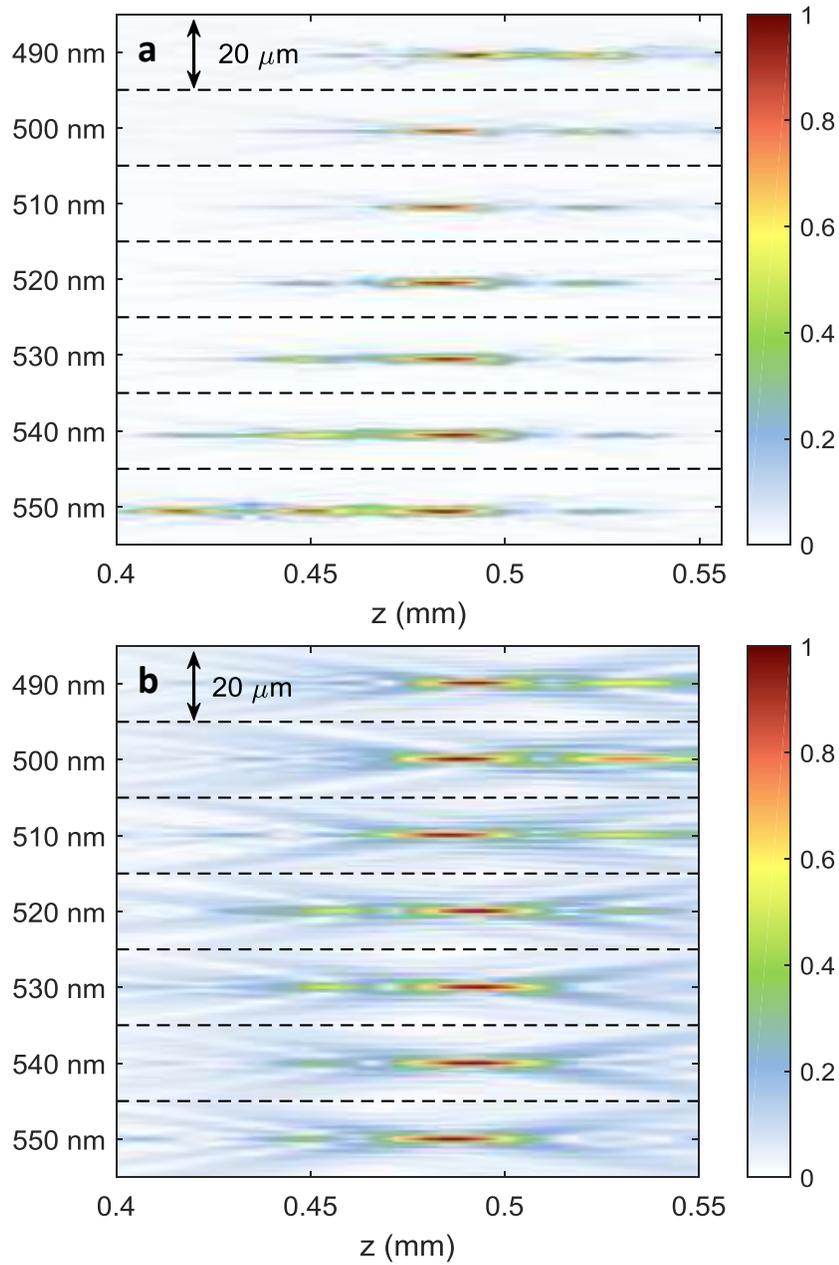

**Figure 4.** (a) Measured intensity profiles of the reflected beam by the achromatic metalens (AML) in the *xz*-plane at different wavelengths. Wavelength of the incident light is noted in the left side of the plot. Focal spots of the AML at the different wavelengths appear at the same *z* distance from the AML. The variation of focal length across the wavelength range of 490 nm- 550 nm is negligible with a standard deviation of 2.7 μm. (b) Simulated intensity profiles of the reflected beam by the achromatic metalens (AML) in the *xz*-plane at different wavelengths.

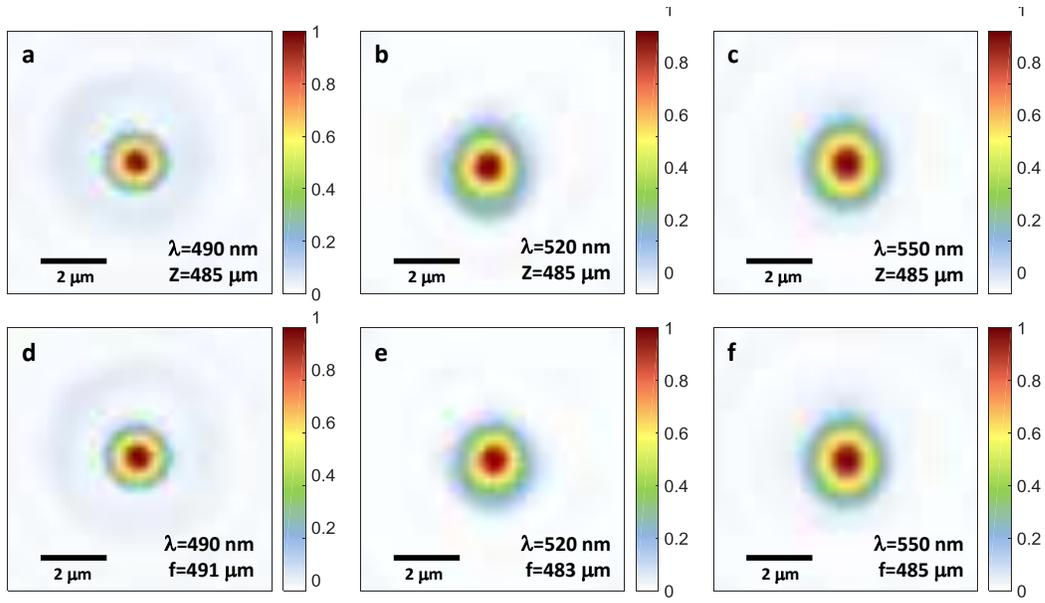

**Figure 5.** Top row: intensity profiles of the achromatic metalens (AML) at a fixed position of z=485 μm at three wavelengths of (a) 490 nm, (b) 520 nm, and (c) 550 nm. Bottom row: focal spots of the AML at three wavelengths of (d) 490 nm, (e) 520 nm, and (f) 550 nm. The vertical (horizontal) cuts of focal spots have a full-width at half-maximum (FWHM) of 1.27 μm (1.29 μm), 1.52 μm (1.45 μm), and 1.68 μm (1.59 μm) at wavelength of 490 nm, 520 nm and 550 nm, respectively. Values of FWHMs from vertical and horizontal cuts are very close revealing the symmetry of the focal spots.

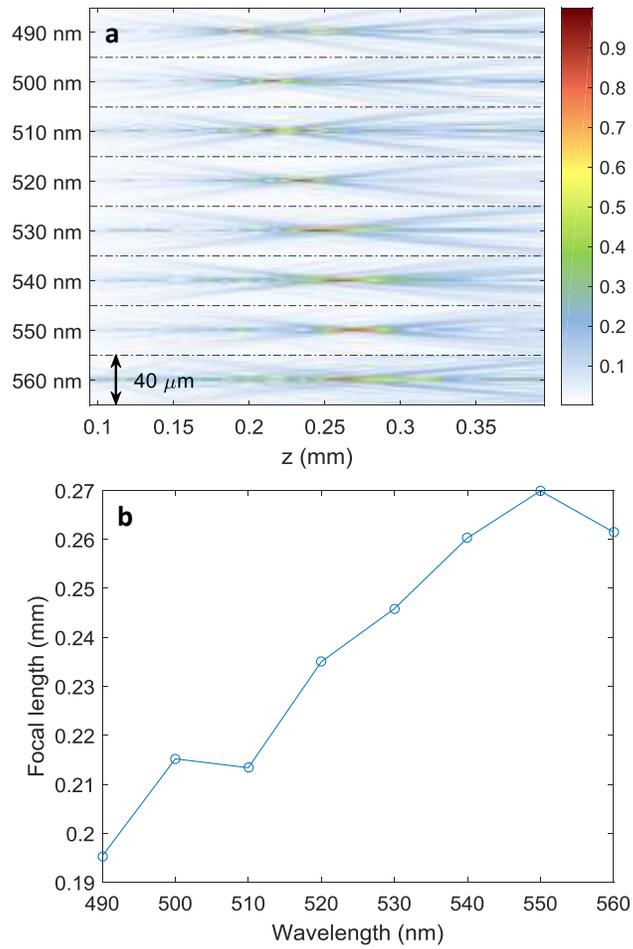

**Figure 6.** (a) Simulated intensity profiles of the reflected beam of a metalens with reverse chromatic dispersion in the *xz*-plane at different wavelengths. Wavelength of the incident light is noted in the left side of the plot. The metalens is designed to achieve reverse chromatic dispersion. This is evident in (b), where the focal length increases with wavelength.

19